\documentclass[preprint,showpacs,prc,aps,showkeys]{revtex4-1}
\usepackage{graphicx}
\usepackage{dcolumn}
\usepackage{bm}
\usepackage{hyperref}

\begin{document}

\preprint{APS/123-QED}

\title{Possibility of Antimagnetic Rotation in odd-A Cd isotopes }

\author{Santosh Roy
}\altaffiliation[Also at ]{S. N. Bose National Centre for Basic Sciences. Block JD, Sector III, Saltlake City, Kolkata 700098, India}
\author{ S. Chattopadhyay}
\affiliation{Saha Institute of Nuclear Physics, 1/AF Bidhannager Kolkata, 700 064, India}

\date{\today}

\begin{abstract}
The classical particle plus rotor model for antimagnetic rotation (AMR)
has been used to investigate the possibility of observation of AMR in the
high spin levels of $^{105,\,107,\,109}$Cd. The calculated I($\omega$) plot
and B(E2) values have been compared with the available experimental data.
\begin{description}

\item[PACS numbers] 21.10.Re, 21.10.Tg, 21.60.Ev, 23.20.-g, 27.60.+j

\end{description}
\end{abstract}

\keywords{AntiMagnetic Rotation, classical particle rotor model,odd-A Cd}
\maketitle

\section{Introduction}
\indent The different manifestations of Shears mechanism has been found
in Cd-isotopes, namely M1 band ($^{110}$Cd)~\cite{rmc1}, band crossing in M1
 band ($^{108}$Cd)~\cite{santoshPRC} and antimagnetic rotation(AMR)
($^{106,\, 108,\,110}$Cd)~\cite{simons, pd1, PLBsantosh}. 
All these observed features have been well described by the geometrical model 
of Shears mechanism.

\indent Recently a systematic study of antimagnetic rotation have been carried
out for the even-even Cd-isotopes, namely $^{106,\,108,\,110}$Cd within the
framework of classical particle rotor model~\cite{PLBsantosh}. In this work,
observed I($\omega$) plots and the measured B(E2) values for all the three
isotopes have been well reproduced by this model calculations.

\begin{figure}
\includegraphics*[scale=0.7]{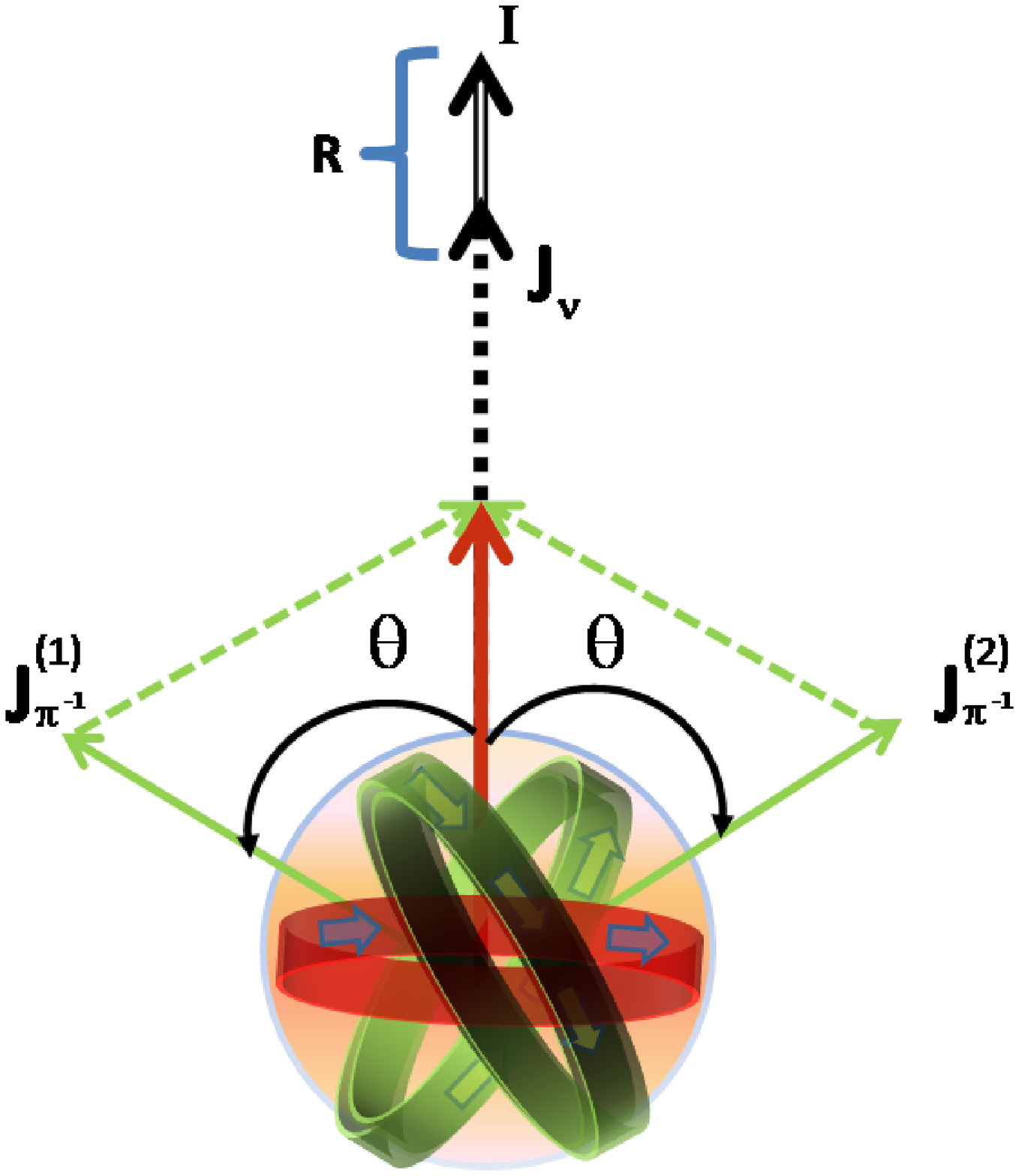}
\caption{\label{fig:fig1}The angular momentum vector diagram for AMR+rotation scenario where {\bf I, R,} ${\textbf j_\pi}$ and ${\textbf j_\nu}$ are the total, rotational, proton hole and neutron particle angular momenta.}
\end{figure}

\indent The vector diagram for the particle and hole angular momenta 
coupling scheme for AMR is shown in Fig.~\ref{fig:fig1} which corresponds
to a symmetric double shear structure. For Cd isotopes, there are two
proton holes in the $g_{\frac{9}{2}}$ orbitals whose angular momenta 
(\textbf{$\bm{j}_\pi$}) are along the symmetry axis, while the angular
momentum of the neutron (\textbf{$\bm{j}_\nu$}) particles in 
$h_{\frac{11}{2}}$/$g_{\frac{7}{2}}$/$d_{\frac{5}{2}}$ orbitals are along
the rotational axis~\cite{simons, pd1} . Thus, $j_\pi=\frac{9}{2}$  and $j_\nu=a*j_\pi$,
where `a' is the ratio of the magnitude of proton and neutron angular momentum
for a specific single particle configuration.
The only important degree of freedom for this model is the angle between
\textbf{$\bm{j}_\pi$} and \textbf{$\bm{j}_\nu$} and is known as the shears
angle ($\theta$). Due to the symmetry of the double shear, the angle between
the two hole vectors is $2\theta$. The interactions between these particle-hole
and the hole-hole blades can be modeled as $P_2$-type forces~\cite{machia1}. It has been
argued by Macchiavelli et. al., that such an interaction may be mediated through
the core by a particle-vibrational coupling involving quadrupole phonon~\cite{machia2}.
The systematic study of AMR in even-even Cd-isotopes indicates that the
strengths of these particle-hole and hole-hole interactions are 1.2 MeV and 
$0.15-0.2$ MeV, respectively, for this mass region~\cite{PLBsantosh}.

\indent In the present work, a systematic theoretical study has been presented
for $^{105,\,107,\,109}$Cd and the calculated I($\omega$) plot and measured B(E2) values have been
compared with the available experimental data.

\section{Classical Particle-rotor Model for AMR}

\indent In this model the energy E(I) is given by~\cite{PLBsantosh}, 
\begin{equation}
E(I)=\ \frac{(\bm{I-j_{\pi} - j_{\nu}})^2}{2\Im}+\ \frac{V_{\pi \nu}}{2}(\frac{3 {\cos^2 \theta}-1}{2})\ 
+\  \frac{V_{\pi \nu}}{2}(\frac{3 {\cos^2 (-\theta)}-1}{2}) - \ \frac{V_{\pi \pi}}{n} (\frac{3 {\cos^2 (2\theta)}-3}{2})
\label{eqn:eqn1}
\end{equation}

\noindent where the first term is the rotational contribution and the rest of 
the terms are the shear contributions, $V_{\pi \nu}$ = 1.2 MeV and 
$V_{\pi \pi}$ = 0.2 MeV. `n' is the scaling factor between $V_{\pi \nu}$ and
$V_{\pi \pi}$ and is determined by the actual number of particle-hole pairs for
a given single-particle configuration. 

\indent The corresponding total angular momentum can be evaluated by
imposing the energy minimization condition as a function of $\theta$
and is given by,
\begin{equation}
I=\ aj+\ 2j \cos\theta+\frac{1.5 \Im V_{\pi\nu} \cos \theta}{j}-\frac{6 \Im V_{\pi \pi} \cos 2\theta\  \cos\theta}{nj}
\label{eqn:eqn2}
\end{equation}

The first two terms represents the contribution from shears mechanism 
($I_{sh}$) as is evident from Fig.~\ref{fig:fig1}. At the band head 
($\theta=90^\circ$), I = $j_\nu$ = aj, which corresponds to the aligned angular
momentum of the neutrons due to core rotation. This implies that there is
a band head frequency which corresponds to the alignment frequency of 
the neutrons which is essential for the formation of the shear structure.
The higher momentum states are formed by gradual closing of the shears angle
 and the maximum angular momentum (${I_{sh}}^{max}$) that can be
generated through AMR due to complete alignment of the two proton holes 
($\theta=0^\circ$) in $g_{9/2}$ orbital is
\begin{equation}
{I_{sh}}^{max}=j_\nu+\frac{9}{2}+\frac{7}{2}
\label{eqn:eqn3}
\end{equation}

\indent The significance of the third and fourth terms of Eq.~\ref{eqn:eqn2}
becomes apparent if we determine the expression for the frequency associated
with the shears mechanism ($\omega_{sh}$). This can be computed through
$\left(\frac{dE_{sh}}{d\theta}\right)/\left(\frac{dI_{sh}}{d\theta}\right)$ and is given by,
\begin{equation}
\omega_{sh}=\ (1.5 V_{\pi\nu}/j)\cos\theta -(6V_{\pi\pi}/nj)\cos2\theta \ \cos\theta
\label{eqn:eqn4}
\end{equation}

\noindent Thus, the third and fourth terms of the Eq.~\ref{eqn:eqn2} are
equal to the product of rotational moment of inertia {$\Im$} and shears
frequency ($\omega_{sh}$). They represent the interplay between
collective and shears mechanism and Eq.~\ref{eqn:eqn2} can be 
re-written as
\begin{equation}
I=I_{sh}+\Im \omega_{sh}
\label{eqn:eqn5}
\end{equation}

\indent It is to be noted that the magnitude of $\Im$ determines the
extent of the interplay in generation of angular momentum in AMR+rotation
model. This value can be estimated from Eq.~\ref{eqn:eqn5} 
\begin{equation}
\Im \omega_{sh}|_{(\theta=0^\circ)}=I_{max}-{I_{sh}}^{max}
\label{eqn:eqn6}
\end{equation}

where, $I_{max}$ is the highest observed angular momentum state, 
${I_{sh}}^{max}$ is given by Eq.~\ref{eqn:eqn3} and 
\begin{equation}
\omega_{sh}|_{\theta=0^\circ} =
\left(\frac{1.5 V_{\pi \nu}}{j}\right)-\left(\frac{6 V_{\pi \pi}}{nj}\right)
\label{eqn:eqn7}
\end{equation}

In case of AMR+rotation model, the rotational frequency ($\omega$) is given by
\begin{equation}
\omega=\omega_{rot}-\omega_{sh}
\label{eqn:eqn8}
\end{equation}

where $\omega_{rot}=\frac{1}{2\Im_{rot}}(2I+1)$ is the core rotational 
frequency and $\Im_{rot}$ is the core moment of inertia, whose value can be
estimated from the slope of the $I(\omega)$ plot for the ground state band 
(before the neutron alignment). The relative negative sign in 
Eq.~\ref{eqn:eqn7} indicates that a given angular momentum state for AMR+rotation 
will be formed at a lower frequency as compared to that due to
 pure rotation.

\indent Thus, all the parameters of the present model can either be fixed
from experimental data or from the systematics of the mass region. Using
Eq.~\ref{eqn:eqn5} and Eq.~\ref{eqn:eqn8}, the theoretical $I(\omega)$
plots have been calculated for $^{105,\,107,\,109}$Cd with 
$V_{\pi \nu}=1.2 $ MeV and $V_{\pi \pi}=0.2$ MeV.

\indent The B(E2) values have been calculated following the expression~\cite{rmc2}
\begin{equation}
B(E2)=\ \frac{15}{32 \pi} (eQ_{eff})^2\ {sin^4  \theta}
\label{eqn:eqn9}
\end{equation}

where $eQ_{eff}\sim 1.1$ eb can be used for all the Cd-isotopes since
they have similar deformation and value of $eQ_{eff}$ is almost solely
determined by the two proton holes~\cite{pd1}. It has been 
demonstrated in the case of $^{110}$Cd~\cite{PLBsantosh}, that the B(E2) 
values fall slowly in case of AMR+rotation as compared to pure AMR as 
the shear angle closes from $90^\circ$ to $0^\circ$ over a larger 
angular momentum range. 

\section{Results}
\subsection{$^{109}$Cd}

\begin{figure}
\includegraphics*[scale=0.7]{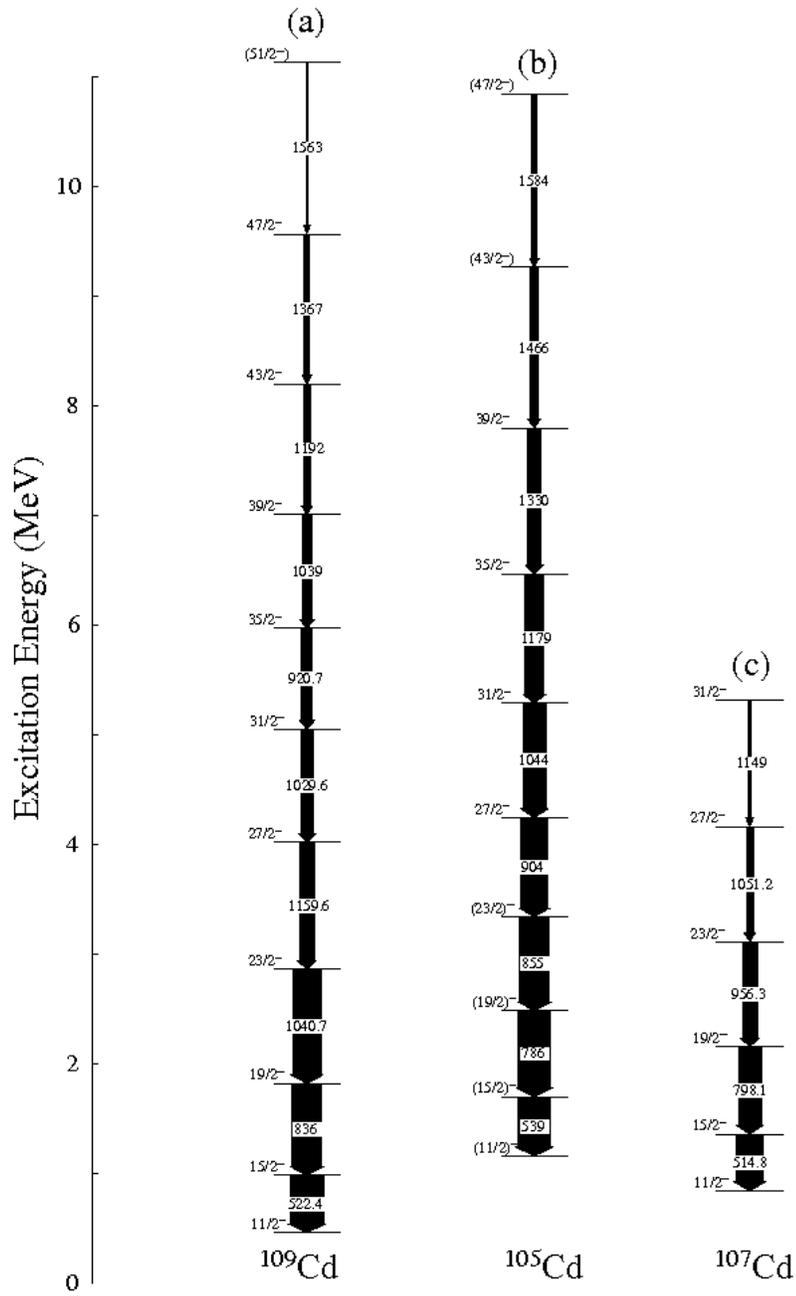}
\caption{\label{fig:fig2}~ Partial level schemes of the negative parity yrast bands of (a) $^{109}$Cd~\cite{chiara}, (b) $^{105}$Cd~\cite{jerrestam1} and  (c) $^{107}$Cd~\cite{jerrestam2}.}
\end{figure}

The negative parity yrast sequence of $^{109}$Cd~\cite{chiara} is shown in 
Fig.~\ref{fig:fig2}(a). The high spin levels beyond ${31/2}^-$
originates due to $(\pi {g_{(9/2)}}^{-2}) \otimes \nu (h_{11/2})^3$ 
configuration. For these levels the measured $\Im^{(2)}$/B(E2) ratio was
found to be around 165 $\hbar^2$ MeV$^{-1}$ $(eb)^{-2}$ which is an order of
magnitude larger than that expected for a well deformed rotor. Thus, it was
concluded that these levels originates due to AMR~\cite{chiara}.

\begin{figure}
\includegraphics*[scale=0.7]{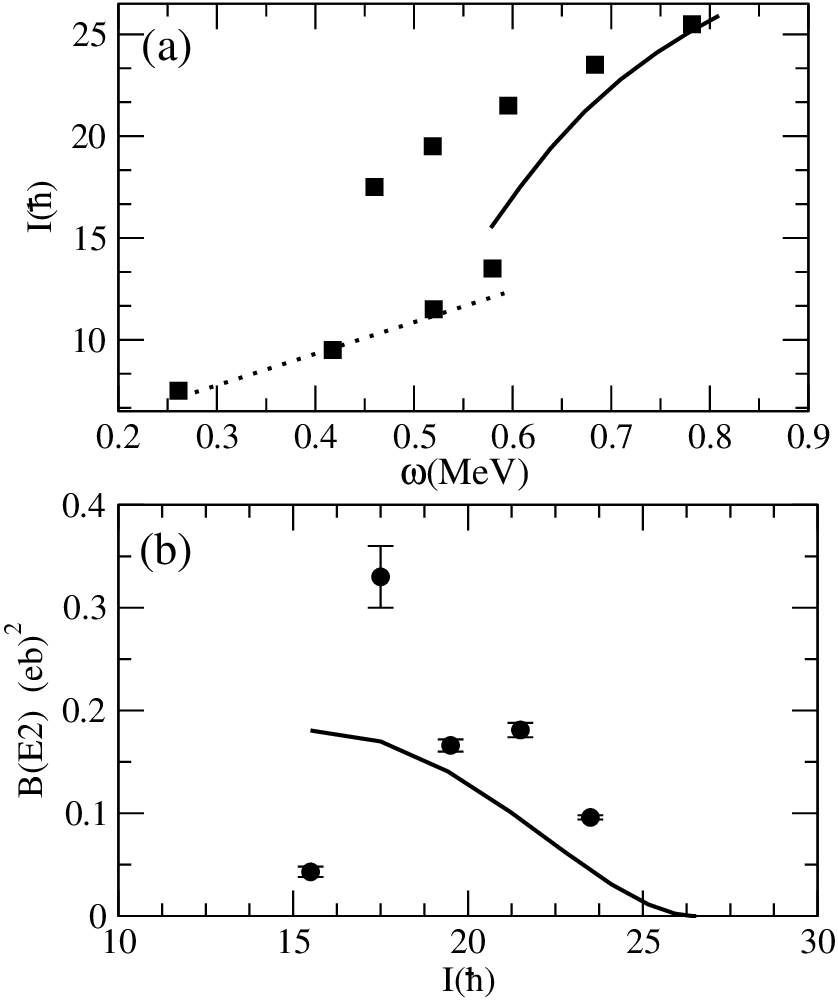}
\caption{\label{fig:fig3}~ The observed I($\omega$) plots (a) and B(E2) values (b) in $^{109}$Cd. The solid line represents the calculated values using the classical particle rotor model for $V_{\pi \nu}$ = 1.2 MeV,
 $V_{\pi \pi}$ = 0.2 MeV, $\Im\sim6$ MeV$^{-1}$ $\hbar^2$  and $\Im_{rot}\sim15$ MeV$^{-1}$ $\hbar^2$. The dashed line in (a) represents the I($\omega$) plot for a rotor with moment of inertia 15  MeV$^{-1}$ $\hbar^2$ shifted by $\sim 4 \hbar$ along the y-axis.}
\end{figure}

\indent The AMR band head spin ($j_\nu$) for this configuration is
27/2 $\hbar$ (11/2$\hbar$ + 9/2$\hbar$ + 7/2$\hbar$) which is shifted 
by $2\hbar$ due to core
rotation. A similar situation has been observed in 
$^{106}$Cd~\cite{simons}. The double shear structure is formed by the 
three aligned neutrons and the two proton holes. Thus, there are six 
possible particle-hole pairs and one hole-hole pair which indicates
n = 6 in Eq.~\ref{eqn:eqn1}. This band has been established up to 51/2 
$\hbar$ ($I_{max}$) while ${I_{sh}}^{max}=47/2 \ \hbar$. Thus, 
$\Im \omega_sh|_{(\theta=0^\circ)}=2 \ \hbar$ which gives 
$\Im=6$ MeV$^{-1}$ $\hbar^2$ from Eq.~\ref{eqn:eqn6}. 
$\Im_{rot}$ has been found to be 15 MeV$^{-1}$ $\hbar^2$, which is 
the slope of the I($\omega$) plot for the ground state band. This is shown
in Fig.~\ref{fig:fig3}(a) by the dotted line that has been shifted by
$\sim4 \hbar$ which corresponds to the aligned angular momentum of the 
valance neutron of $^{109}$Cd. 

\indent With this set of fixed parameters, the I($\omega$) plot for
the neutron aligned yrast band of $^{109}$Cd has been calculated and
shown as the solid line in Fig.~\ref{fig:fig3}(a). The calculated 
frequencies have been shifted by the band head frequency of 0.51 MeV.
It is evident from the figure that the AMR configuration is non-yrast
over the entire frequency range. The calculated B(E2) values are shown
by the solid line in Fig~\ref{fig:fig3}(b). It is apparent that the
measured B(E2) values are about twice as large as that predicted by the 
model. 

\indent Thus, the present calculations seem to indicate that the high
spin yrast levels of $^{109}$Cd do not originate due to AMR.

\subsection{$^{105}$Cd}

The negative parity yrast sequence of $^{105}$Cd~\cite{jerrestam1} has been 
established upto 47/2$^-$ and is shown in 
Fig.~\ref{fig:fig2}(b). The levels beyond 23/2$^-$ originate due to
$\pi {g_{9/2}}^{-2}$ $\otimes$ $\nu [h_{11/2}, (g_{7/2}/d_{5/2})^2]$
configuration and the lifetimes of these levels have not been 
measured~\cite{jerrestam1}. For this configuration n = 6 and the AMR band head
is 23/2 $\hbar$ [11/2 $\hbar$ + 7/2 $\hbar$ +5/2 $\hbar$]. This implies
that ${I_{sh}}^{max}=39/2 \ \hbar$. Thus, for $^{105}$Cd, 
$\Im \omega_{sh}|_{(\theta=0^\circ)}= 4\, \hbar$. This leads to $\Im=11.2$
MeV$^{-1}$ $\hbar^2$. The value for $\Im_{rot}$ has been found to be 14 
MeV$^{-1}$ $\hbar^2$ in the same way as described for $^{109}$Cd.

\indent The experimental and calculated I($\omega$) plot for the neutron
aligned band have been shown in Fig.~\ref{fig:fig6}(a), where the 
calculated frequencies have been shifted by the band head frequency 
($\sim$0.4 MeV). It is interesting to note that the AMR configuration is
 non-yrast at lower frequencies but becomes yrast around 
$\hbar\omega$= 0.60 MeV which corresponds to 35/2 $\hbar$ level. 
Thus, the levels beyond 31/2 $\hbar$ are expected to originate due to AMR.
Fig.~\ref{fig:fig6}(b) shows the predicted B(E2) values over the entire
frequency range. The predicted B(E2) values for 31/2 $\hbar$ 
($\theta= 73.5^\circ$)
, 35/2 $\hbar$ ($\theta= 64.5^\circ$) and  39/2 $\hbar$ ($\theta=54^\circ$)
 levels are 0.15, 0.12 and 0.07 $(eb)^2$, respectively.

\indent The present model predicts that the levels beyond 31/2 
$\hbar$ originates due to an interplay between AMR and collective 
rotation. This prediction can be established by measuring the B(E2)
 transition rates for the levels beyond 31/2 $\hbar$ and comparing them
with the predicted values.

\subsection{$^{107}$Cd}

\indent The negative parity yrast sequence of $^{107}$Cd~\cite{jerrestam2} 
is known only
upto 31/2 $\hbar$ (shown in Fig.~\ref{fig:fig2}(c)), which needs to be extended in
 order to investigate the possibility of AMR in $^{107}$Cd. The aligned
angular momenta ($i$) of $^{105,\, 107}$Cd has been plotted against the
rotational frequency $(\hbar \omega)$ in Fig. 5. It is apparent from the 
figure that the alignment gain is similar in the two cases and is around
$4\hbar$. Thus, the neutron aligned yrast configuration for $^{107}$Cd
is also expected to be $(\pi {g_{9/2}}^{-2})$ $\otimes$
$[h_{11/2}, (g_{7/2}/d_{5/2})^2]$, which is same as that of $^{105}$Cd.

\indent Since this band has not been observed upto the highest spin, the
I($\omega$) plot has been calculated for three possible values of
$\Im\omega_{sh}|_{(\theta=0^\circ)}$ = $2 \hbar$,  $4 \hbar$ and  $6 \hbar$
and the predicted values are shown in Fig.~\ref{fig:fig6}(a) 
with dashed, solid and 
dot-dashed lines, respectively, for $\Im_{rot}= 14$ MeV$^{-1}$ $\hbar^2$.
Once the higher spin levels of this band is established, 
$\Im \omega_{sh}|_{\theta=0^\circ}$ will get fixed as for this band ${I_{sh}}^{max}$
 = $39/2 \hbar$. Thus, the model calculations will have a definitive prediction of
$I(\omega)$ behavior of the band if it originates due to AMR+rotation.

\indent The predicted B(E2) values are shown in Fig.6 (b) for $\Im\omega_{sh}|_{(\theta=0^\circ)}$.
 A comparison of the experimental values with these predicted values will
 be an additional check to establish whether the high spin levels of
$^{107}$Cd originate due to AMR+rotation.

\section{Conclusion}

\indent The classical particle rotor model has been applied for 
three odd-A isotopes of Cd, namely $^{105,\,107,\,109}$Cd,
 where AMR is expected since this excitation mode has been reported in their
 even-even partners. A comparison of I($\omega$) plot and B(E2) values 
for $^{109}$Cd seems to indicate that the high spin levels do not originate 
due to Shears mechanism. On the other hand, levels beyond 31/2 $\hbar$ 
in $^{105}$Cd is predicted to originate due to an interplay of 
antimagnetic and collective rotation. However this prediction can only 
be established through the measurement of B(E2) transition rates for these 
levels. In case of $^{107}$Cd, the negative parity yrast band have not 
been established. Thus, model dependent predictions for I($\omega$) plot and B(E2) 
values have been presented. 

\indent These calculations complement those performed for even-even Cd isotopes 
within a common framework of a classical particle rotor model where $V_{\pi \pi}$ = 1.2 MeV 
and $V_{\pi \nu}$ =0.2 MeV have been used for all the Cd-isotopes, namely $^{105-110}$Cd.

\newpage

\begin{figure}
\includegraphics*[scale=0.7]{New105Cd.eps}
\caption{\label{fig:fig4}~ The observed I($\omega$) plots (a) and B(E2) values (b) in $^{105}$Cd. The solid line represents the calculated values using the classical particle rotor model for $V_{\pi \nu}$ = 1.2 MeV, $V_{\pi \pi}$ = 0.2 MeV, $\Im\sim11$ MeV$^{-1}$ $\hbar^2$  and $\Im_{rot}\sim15$ MeV$^{-1}$ $\hbar^2$. The dashed line in (a) represents the I($\omega$) plot for a rotor with moment of inertia 15  MeV$^{-1}$ $\hbar^2$ shifted by $\sim 4 \hbar$ along the y-axis.}
\end{figure}

\begin{figure}
\includegraphics*[scale=0.7]{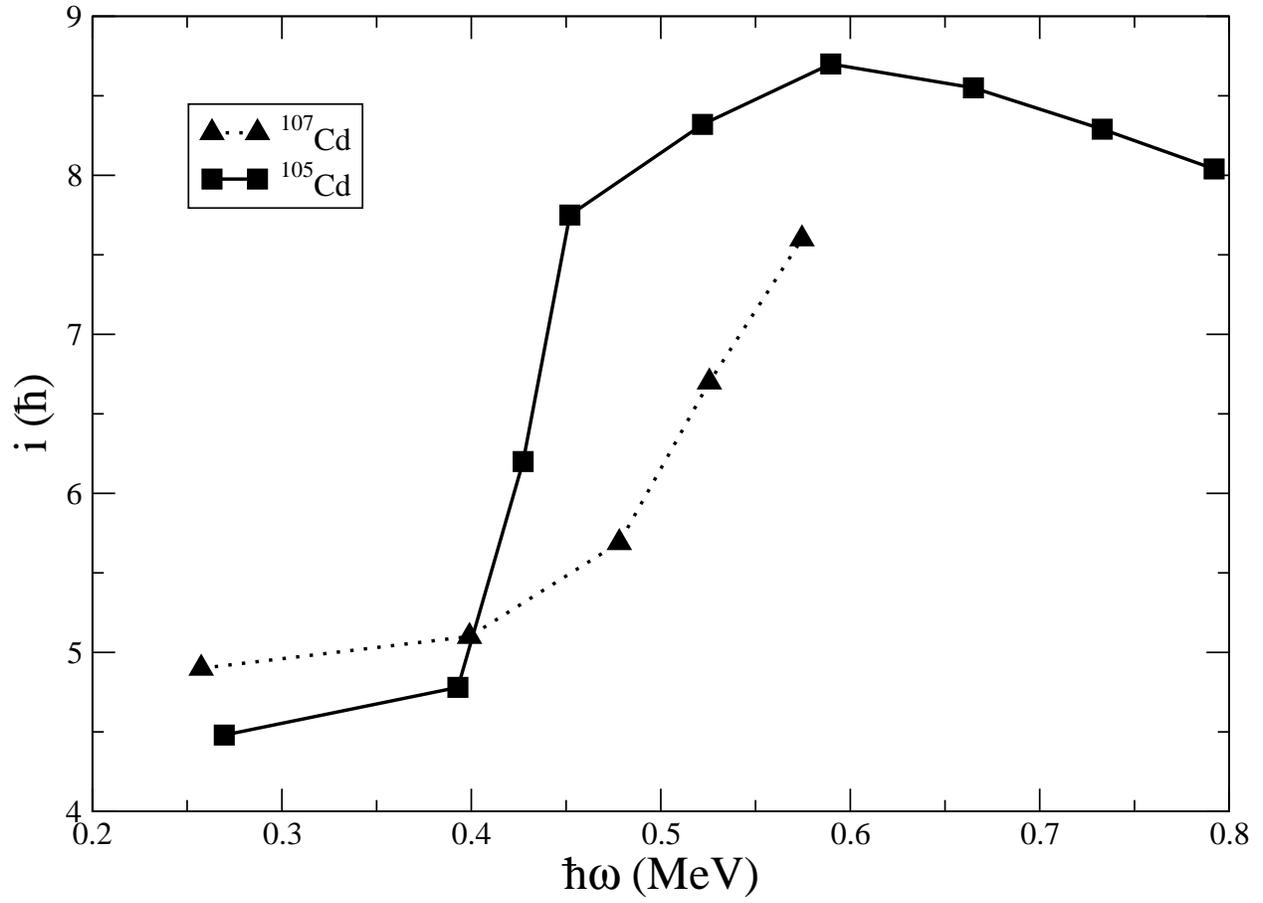}
\caption{\label{fig:fig5}~Experimantal aligned angular momentum for negative parity yrast band in $^{105}$Cd~\cite{jerrestam1} (solid line) and 
$^{107}$Cd~\cite{jerrestam2} (dotted line). }
\end{figure}

\begin{figure}
\includegraphics*[scale=0.7]{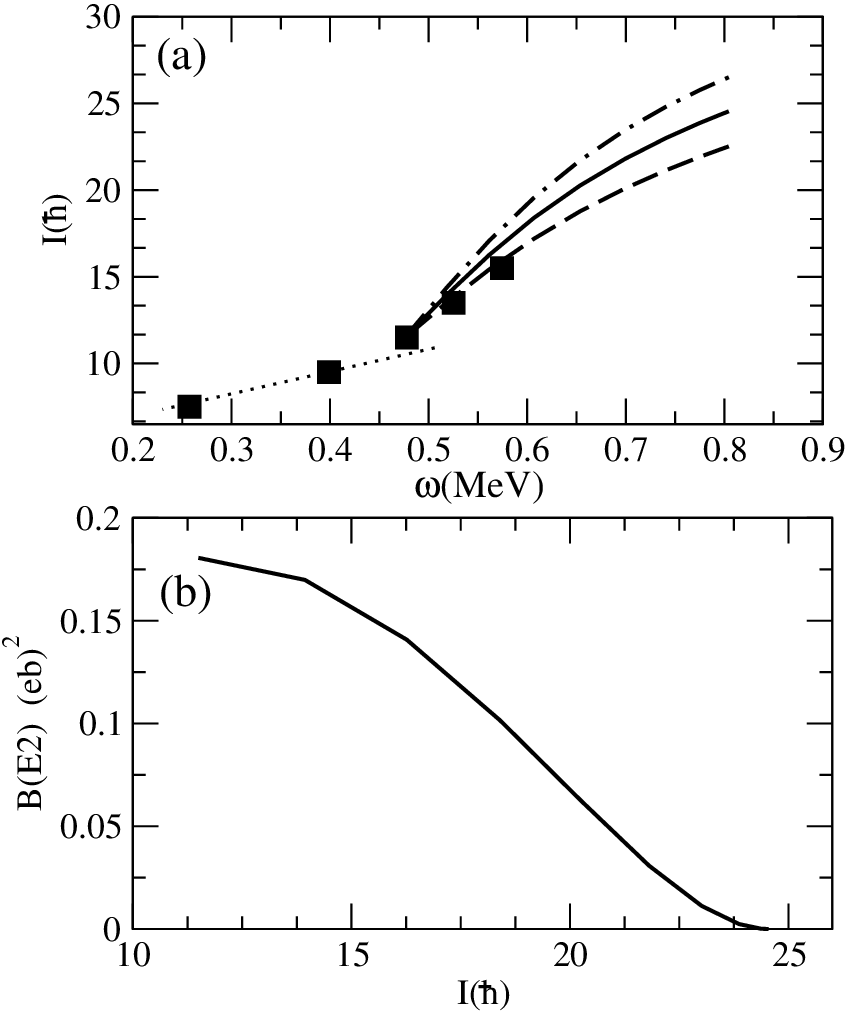}
\caption{\label{fig:fig6}~ The observed I($\omega$) plots (a) and B(E2) values (b) in $^{107}$Cd. The dashed, solid and dot-dashed lines represents the calculated values for $\Im\sim 6$ MeV$^{-1}$ $\hbar^2$, $\Im\sim 11$ MeV$^{-1}$ $\hbar^2$ and $\Im\sim 17$ MeV$^{-1}$ $\hbar^2$, respectively. The
other fixed parameters are $V_{\pi \nu}$ = 1.2 MeV, $V_{\pi \pi}$= 0.2 MeV and
$\Im_o=14$ MeV$^{-1}$ $\hbar^2$. The dotted line in (a) represents the I($\omega$) plot for a rotor with moment of inertia 15  MeV$^{-1}$ $\hbar^2$ shifted by $\sim 4 \hbar$ along the y-axis.}
\end{figure}


\newpage


\begin{thebibliography}{99}
\bibitem{rmc1} R.M. Clark {\sl et al.}, Phys. Rev. Lett. {\bf 82}, 3220 (1999).
\bibitem{santoshPRC} Santosh Roy {\sl et al.},  Phys. Rev. C {\bf 81}, 054311 (2010)
\bibitem{simons}A. J. Simons {\sl et al.}, Phys. Rev. Lett. {\bf 91}, 162501 (2003).
\bibitem{pd1}P. Datta {\sl et al.}, Phys. Rev. C {\bf 71}, 041305(R) (2005).
\bibitem{PLBsantosh} Santosh Roy {\sl et al.}, Physics Letters B (2010), doi:10.1016/j.physletb.2010.10.018
\bibitem{machia1} A.O. Macchiavelli {\sl et al.}, Phys. Rev. C {\bf 57}, R1073 (1998).
\bibitem{machia2} A.O. Macchiavelli {\sl et al.},Phys. Rev. C {\bf 58}, R621–R623 (1998).
\bibitem{sugawara} M. Sugawara {\sl et al.}, Phys. Rev. C {\bf 79}, 064321 (2009).
\bibitem{rmc2} R.M. Clark and A. O. Macchiavelli, Annu. Rev. Nucl. Part. Sci. {\bf 50}, 1 (2000).
\bibitem{chiara}C. J. Chiara {\sl et al.}, Phys. Rev. C {\bf 61}, 034318 (2000).
\bibitem{jerrestam1} D. Jerrestam {\sl et al.}, Nucl. Phys. {\bf A593}, 162 (1992).
\bibitem{jerrestam2} Dan Jerrestam {\sl et al.}, Nucl. Phys. {\bf A545}, 835 (1992).

\end{thebibliography}
\end{document}